%% file: main.tex
\documentclass[10pt,conference,a4paper]{IEEEtran}

\usepackage[utf8]{inputenc}

\usepackage{blindtext, graphicx}
\usepackage[utf8]{inputenc}
\usepackage[english]{babel}
\usepackage{csquotes}
\usepackage{wrapfig}
\usepackage{float}
\usepackage{verbatim}
\usepackage{subcaption}
\usepackage{import}
\usepackage{dirtytalk}
\usepackage{subfiles}
\usepackage{array}
\usepackage{multirow}

\usepackage{lscape}
\usepackage{algorithm}
\usepackage{algorithmic}
\usepackage{rotating}
\usepackage[table,xcdraw]{xcolor}

\usepackage{times}

\DeclareGraphicsExtensions{.png,.eps,.ps,.pdf}

\usepackage{url}
\usepackage{hyperref}

\usepackage[most]{tcolorbox}

\usepackage{multirow}
\usepackage{tabularx}
\usepackage{makecell}
\usepackage{svg}

\usepackage{pifont}
\definecolor{ao}{rgb}{0.0, 0.5, 0.0}
\definecolor{amber}{rgb}{1.0, 0.49, 0.0}

\let\labelindent\relax
\usepackage{enumitem}

\usepackage{lscape} 

\graphicspath{{images/}}

\usepackage{tikz} 
\newcommand{\orcidicon}{\includegraphics[width=0.32cm]{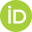}}
\foreach \x in {A, ..., Z}{%
\expandafter\xdef\csname orcid\x\endcsname{\noexpand\href{https://orcid.org/\csname orcidauthor\x\endcsname}{\noexpand\orcidicon}}
}

\begin{document}
%

\title{Toward a Military Smart Cyber Situational Awareness (CSA)}

\author{\IEEEauthorblockN{
\orcidA{}Anthony Feij\'oo-Añazco$^1$,
\orcidB{}Antonio L\'opez Mart\'inez$^2$,
\orcidC{}Daniel D\'iaz-L\'opez$^{1,2}$\\
\orcidD{}Angel Luis Perales G\'omez$^{3}$,
\orcidF{}Pantaleone Nespoli$^{1}$,
\orcidE{}Gregorio Mart\'inez P\'erez$^{1}$
}
\IEEEauthorblockA{$^1$Department of Information and Communications Engineering, University of Murcia, 30100, Murcia, Spain\\
\{anthonyjosep.feijooa, antonio.lopezm2, danielorlando.diaz, pantaleone.nespoli, gregorio\}@um.es}
\IEEEauthorblockA{$^2$School of Engineering, Science and Technology, Universidad del Rosario, Bogot\'a, Colombia\\
danielo.diaz@urosario.edu.co}
\IEEEauthorblockA{$^3$Department of Computers Engineering and Technology, University of Murcia, 30100, Murcia, Spain\\
angelluis.perales@um.es}

}


\maketitle

\begin{abstract} 
The development of technology across multiple sectors and the growing importance of cyber warfare make the development of Cyber Situational Awareness (CSA) a fundamental component of any cyber defense strategy. CSA, as a practice, enables understanding of the current landscape within an organization or critical infrastructure, anticipating potential threats, and responding appropriately to cyber risks. With CSA, we are not simply seeking a passive point of view, but rather informed decision-making that allows us to improve response times and monitor the consequences and effects an attack has on one of our elements and how it will affect other elements it interacts with. In this paper, we review 5 CSA platforms, seeking differentiating characteristics between each proposal and outlining 6 proposed criteria that can be applied when creating a military smart CSA platform. To this end, we have validated the proposed criteria in CRUSOE, an open-source CSA platform developed by CSIRT-MU. After applying some modifications and experiments, it turned out to be applicable to this field.
\end{abstract}


{\bf Contribution type:}  {\it  Original research }

\input{Introduction}

\input{StateOfTheArt.tex}
\input{Comparison.tex}
\input{Proposal.tex}

\input{Experiments.tex}

\input{Conclusions.tex}

\section*{Acknowledgment}
This work has been co-funded by the European Union (project ECYSAP EYE). Views and opinions expressed are however those of the author(s) only and do not necessarily reflect those of the European Union or the European Defence Fund. Neither the European Union nor the granting authority can be held responsible for them.
This work has also been partially supported by MCIN/AEI/10.13039/501100011033 NextGeneration EU/PRTR, UE, under Grant TED2021-129300B-I00, by MCIN/AEI/10.13039/501100011033/FEDER, UE, under Grant PID2021-122466OB-I00, by  the Spanish National Institute of Cybersecurity (INCIBE) by the Recovery, Transformation and Resilience Plan, Next Generation EU under the strategic project DEFENDER, by the CyberDataLab (Cybersecurity and Data Science Laboratory) at the University of Murcia (Spain), and the School of Engineering, Science and Technology at the University of Rosario (Colombia).

\bibliographystyle{IEEEtran}
\bibliography{bibliography.bib}{}

\end{document}

%% file: Introduction.tex
\section{Introduction}\label{sec:intro} 

Situational Awareness (SA) arises from the military environment where the aim is to know the situation of different elements and estimate how a decision may affect them~\cite{Torvald2023, FRANKE2014}. In the cybersecurity domain, Cyber Situational Awareness (CSA) applies SA principles to cyberspace, focusing on real-time monitoring, analysis, and response to threats~\cite{esteve2016cyber}. Specifically, CSA involves identifying the behavior of adversaries and how their attack may affect assets to prioritize strategies and ensure the successful course of operations~\cite{CyCOP2023}.

CSA plays an important role in cyber defense due to its dynamic and predictive functionality, as well as real-time monitoring, allowing commanders to take proactive measures to allocate resources and validate system designs against threats. This capability uses Mission Engineering (ME), a methodology that aligns technical capabilities with operational objectives through threat impact modeling~\cite{dillabaugh2020cybercop}. ME relies on CSA data to quantify how attacks affect strategic objectives, such as continuity of military operations or essential services, and can transform data into relevant tactical decisions~\cite{Llopis2018}.

In recent years, some technological CSA-devoted platforms have emerged to support the objectives of CSA. Military solutions, like the ones described in~\cite{CyCOP2023, Jiang2022, Machado2014ConceptualAF}, aim to integrate operational data's cyber and physical perspectives in a hybrid decision-making system. In addition, different studies discuss the importance of visually transmitting data to non-expert personnel and analyze how this can be represented~\cite{Skopik2022}.

CSA platforms play a crucial role in military cyber defense by providing commanders with real-time information that can be interacted with. These intelligent platforms are responsible for transforming numerous data streams into concentrated and displayable information that allows officers to understand the operational context, enabling them to make tactical decisions that mitigate risks by anticipating potential risks and ensuring mission continuity. Thus, this paper aims to advance the comprehension of requirements and the need for a military smart CSA platform. This study seeks to find an open-source hybrid tool with cyber defense and asset management capabilities that can be used in both civilian and military environments.

Thus, the main contributions are summarized as follows:
\begin{itemize}
    \vspace*{-0.2cm}
    \item Study of 5 CSA platforms, identifying functional differences, integration capabilities, and architectures, identifying benefits and limitations of each proposal.
    \item Proposal of 6 criteria to be considered in designing a military smart CSA platform considering multi-domains: users, data sources, operational environments, supported use cases, visualization, and military requirements.
    \item Validation of the proposed criteria through the development of a cyber defense scenario deployed in an open-source platform for CSA (i.e., CRUSOE), allowing identification of gaps and opportunities for extension.
    \vspace*{-0.2cm}
\end{itemize}

This paper is structured as follows: Section~\ref{sec:sota} reviews selected papers on existing CSA platforms, analyzing their main contributions. Section~\ref{sec:comparison} compares different CSA platforms, identifying their main features. Section~\ref{sec:proposal} defines criteria required for a military smart CSA platform. Section~\ref{sec:experiments} presents experiments conducted on CRUSOE. Finally, Section~\ref{sec:conclusions} summarizes the conclusions and outlines future work.

%% file: StateOfTheArt.tex
\section{State of the art}\label{sec:sota} 

National defense organizations have long seen the need to address the cyber dimension as another field of operations to achieve dominance. The European Defense Agency (EDA) has highlighted, through its interest in its capability development programs funded by the European Defense Fund (EDF)~\cite{EDF}, the importance of enhancing the cyber resilience of Member States and, more importantly, its specific interest in technological development projects focused around cyber situational awareness and cyber defence operations capabilities. Aligned with this call to increase SA scientific effort, this section details some of the most important works related to CSA.  

As early as 2014, a Systematic Literature Review (SLR) on CSA was published~\cite{FRANKE2014}, identifying CSA research thematic clusters and future research directions. Two key points of this study highlight, firstly, how most CSA research remains theoretical, as just a few studies validate real-world impact using empirical data. Secondly, the lack of a unified framework or toolset indicates that existing studies focus on specific applications but lack holistic integration. 


One of the CSA research thematic clusters identified in this previous work is ``Visualization''. In line with this, Jiang et al.~\cite{Jiang2022} extensively reviews different research works focusing on visualization to enhance CSA. They analyze different CSA proposals in terms of the maturity of visualizations and their relationship with key stakeholders, data sources, and CSA levels supported. This SLR notes some finding in the studied CSA works, such as: proposals primary revolve around technical users ($\sim65\%$), there is a lack of projection capabilities in terms of Endsley's situational awareness theory~\cite{Endsley1995} ($\sim18\%$), and a rigorous and industry-suitable evaluation is not generally applied ($\sim24\%$).

Deviating from SLRs, works such as of Georgi et al.~\cite{Georgi2024} accurately stablishe guidelines and standards to follow in this research field, as it delves into metrics and methodologies used to assess SA, identifies key data sources recommended to enhance CSA, and proposes practical evaluation scenarios to measure the effectiveness of CSA techniques, e.g., APT and Insider Threat scenarios.

    
A solid foundation for SA innovation, but one that exhibits one of the previously identified gaps, is the work of Skopik et al.~\cite{Skopik2022}. Such a paper introduces a Cyber Common Operating Picture (CCOP) framework and data model, designed to be flexible, scalable, and structured for the military domain. A case study tested for the Austrian Ministry of Defense (MoD) is also presented, demonstrating the feasibility of centralized cybersecurity reporting. 

In contrast, the study by Ferreira et al.~\cite{Machado2014ConceptualAF} demonstrates a strong commitment to military applications, based on a CSA architecture that integrates kinetic (physical) and cyber environments to enhance decision making and mission planning. Such a work prioritizes Mission Planning and Cyber Threat Mitigation (projection and action), yet this work shows that further implementation efforts are required instead.

Some of the studied proposals showcase the performance of tools with real development, as is the case of Llopis et al.~\cite{Llopis2018}, in which a comparison is made within the ``Visualization'' aspect of two actual military CSA tool developments (not public). The comparison is of great interest even though the identified challenges for both tools are equally important: lack of integration between cybersecurity and mission-oriented decision-making tools, high cognitive workload for users, and missing risk assessment models.



A notable paper bounded by the 2019 ECYSAP project (European Cyber Situational Awareness Platform) is the one presented by Sotelo et al.~\cite{SOTELOMONGE2021869}, which demonstrates the impact of Tactical Denial of Sustainability (ToDS) (negative effects on logistical or operative assets at the very edge of a military network due to cyberincidents).
This proposal, which reinforces CSA platforms with advanced features, is also accompanied by other ECYSAP-associated work of Rodriguez et al.~\cite{Rodríguez-Bermejo2021} where the effect of stress on CSA tool's operators and decision-makers is studied, recommending the incorporation of tools capable of accounting the effect of stress on the staff during cyber operations in CSA frameworks.

Previous works help identify some major gaps in the scientific effort surrounding SA. Scientific activity in this area is thematically diverse in an effort to explore this domain sufficiently to define a solid foundation for SA. However, this has led to a wide array of proposals of diverse thematic that, at present, remain largely unimplemented. Also, most existing studies focus primarily on the first two levels of the SA Endsley's model, consequently, there is a notable lack of proposals (and, more importantly, tools) with active assistance capabilities for projection-related tasks across all operational dimensions, e.g., mission planning, action execution, situation prediction, and impact analysis.


%% file: Comparison.tex
\section{Functional comparison of CSA platforms}\label{sec:comparison} 


Technological CSA-devoted platforms improve threat detection and transform raw data from multiple sources into operationally relevant knowledge, which is essential for tactical, agile decisions. Thus, this section compares 5 CSA chosen platforms from a functional perspective. The comparative Table~\ref{tab:proposals-comp} summarizes the analysis results.

The comparison centers on 6 key functional aspects of CSA platforms. Below is an explanation of each of these aspects.
\begin{itemize}
    \item \textbf{Users}: The target audience of the tool, i.e., decision makers with less technical knowledge, technical-operational personnel, or non-expert personnel.
    \item \textbf{Data sources}: The data sources received by the CSA platform, whether they come from its own devices or other entities, and whether it is specific defense or attack data.
    \item \textbf{Operational environments}: The proposed environments for deploying the CSA include Cybernetic Information Systems (CIS), which involve only electronic components or ``passive entities'', and Tactical Military settings, which encompass various military assets capable of active participation and response
    \item \textbf{Supported use cases}:  When assessing which scenarios the different platforms can be applied to, the use cases presented help to understand their flexibility and adaptability to different situations and operational needs.
    \item \textbf{Visualization techniques}: The transmission of information is a fundamental feature of CSA platforms, as effective transmission of information is crucial for fast and accurate decision making in complex environments. This criterion evaluates the different ways in which each platform presents and communicates critical information.
    \item \textbf{Military approach}: CSA platforms encompass a wide range of functionality applicable to many uses. This criterion focuses on extracting relevant functionalities that may be specifically useful in the military domain, allowing a more accurate assessment of their value in military strategies and planning and executing missions.
\end{itemize}

\begin{table*}[]
\centering
\begin{tabular}{|p{1.2cm}|p{2.1cm}|l|l|l|l|l|}
\hline
\textbf{Proposal} & \multicolumn{1}{c|}{\textbf{Users}} & \multicolumn{1}{c|}{\textbf{Data Sources}} & \multicolumn{1}{c|}{\textbf{\begin{tabular}[c]{@{}c@{}}Operational \\ environments\end{tabular}}} & \multicolumn{1}{c|}{\textbf{\begin{tabular}[c]{@{}c@{}}Supported \\ use cases\end{tabular}}} & \multicolumn{1}{c|}{\textbf{\begin{tabular}[c]{@{}c@{}}Visualization \\ techniques\end{tabular}}} & \multicolumn{1}{c|}{\textbf{\begin{tabular}[c]{@{}c@{}}Military \\ requirements \\ included\end{tabular}}} \\ \hline
\textbf{Kookjin ~\cite{CyCOP2023}} & 
\begin{tabular}[c]{@{}l@{}} Technical team\end{tabular} & 
\begin{tabular}[c]{@{}l@{}} Own. Blue team\end{tabular} & 
\begin{tabular}[c]{@{}l@{}} Cybernetic (CIS)\end{tabular} & 
\begin{tabular}[c]{@{}l@{}} Georeferencing, cyber \\knowledge graph and \\traffic visualization\end{tabular} & 
\begin{tabular}[c]{@{}l@{}} Small screens,\\ Main map\\ scorecard with \\controls and logs. \end{tabular} & 
\begin{tabular}[c]{@{}l@{}}  Visualization of \\operational nodes \end{tabular} \\ \hline

\textbf{Esteve ~\cite{esteve2016cyber}} & 
\begin{tabular}[c]{@{}l@{}} Strategic\\ Operational \\and Technical team\end{tabular} & 
\begin{tabular}[c]{@{}l@{}} Own.\\Red and \\blue team\end{tabular} & 
\begin{tabular}[c]{@{}l@{}} Cybernetic (CIS)\end{tabular} & 
\begin{tabular}[c]{@{}l@{}} Hybrid situational, \\cyber knowledge \\graph and event \\classification \end{tabular}& 
\begin{tabular}[c]{@{}l@{}} Main map , \\ scorecard with \\3D modeling \\and Virtual reality\end{tabular} & 
\begin{tabular}[c]{@{}l@{}} Risk analysis, \\ visualization of \\operational nodes\end{tabular} \\ \hline

\textbf{Noel \newline ~\cite{Noel2023}} & 
\begin{tabular}[c]{@{}l@{}} Technical-strategic \\team\end{tabular} &
\begin{tabular}[c]{@{}l@{}}Own and \\external \end{tabular} & 
\begin{tabular}[c]{@{}l@{}}Cybernetic (CIS),\\ Tactical Military\end{tabular} & 
\begin{tabular}[c]{@{}l@{}}CySU, mission \\impact assessment,\\ threat detection\end{tabular} & 
\begin{tabular}[c]{@{}l@{}}Main map, \\graph visualizations\\ and status panel\end{tabular} & 
\begin{tabular}[c]{@{}l@{}} Mission dependencies \\ and functions, \\ mission dependencies\end{tabular} \\ \hline
\textbf{Dillabaugh ~\cite{dillabaugh2020cybercop}} &
\begin{tabular}[c]{@{}l@{}} Cybernetic \\Emergency \\ Response Team\end{tabular} &
\begin{tabular}[c]{@{}l@{}}Own. Red and \\blue team\end{tabular} &
\begin{tabular}[c]{@{}l@{}}Cybernetic(CIS) \end{tabular} &
\begin{tabular}[c]{@{}l@{}} Operational node \\visualization, Mission \\Engineering and \\cybernetic graph\end{tabular} & 
\begin{tabular}[c]{@{}l@{}}Main map ,\\ Graphs with relations \\ nodes,\\widget information\end{tabular} & 
\begin{tabular}[c]{@{}l@{}}NATO symbols \\\& mission symbols\end{tabular} \\ \hline

\textbf{Husak ~\cite{HUSAK2022}} &
\begin{tabular}[c]{@{}l@{}}Technical team\end{tabular}&
\begin{tabular}[c]{@{}l@{}}Own\end{tabular}&
\begin{tabular}[c]{@{}l@{}}Cybernetic(CIS)\end{tabular}&
\begin{tabular}[c]{@{}l@{}}Incident handling \\ decision support \\ CySA\end{tabular}& 
\begin{tabular}[c]{@{}l@{}}Cyber knowledge graph,\\ mission enginering\\ network \\defense management\end{tabular}& 
\begin{tabular}[c]{@{}l@{}}Mission modeling\\ fingerprinting\\decision support\end{tabular}  \\ \hline
\end{tabular}
\caption{Comparison of different CSA platforms}
\label{tab:proposals-comp}
\end{table*}

\vspace*{-0.1cm}
\subsection{Proposal of Kookjin et al.}

Kookjin et al.~\cite{CyCOP2023} underscored the importance of visualization and the human factor, proposing the design of a Cyber Common Operational Picture (CyCOP) that focuses on analyzing cyberspace operational nodes and aspects of UI. This work establishes a relationship between military mission planning, i.e., JCOPP (Cyberspace Operational Planning Process) \& JTC (Joint Targeting Cycle), and the essential capabilities of an effective COP. However, their Proof of Concept (PoC) only implements stages for defining the operational environment and visualizing threats and assets in the hybrid space, leaving projection-related capabilities for future work.

\vspace*{-0.1cm}
\subsection{Proposal of Esteve et al.}


Esteve et al.~\cite{esteve2016cyber} developed a hybrid model that georeferences cyber domain information from physical assets. Their CyCOP platform features an immersive visual representation of real-time elements and events, showing how threats affect multiple environmental components. The interface supports external data sources and allows manual user input through logical rules, though the specific implementation details are only named rather than explained.

\vspace*{-0.1cm}
\subsection{Proposal of Noel et al.}

Noel et al.~\cite{Noel2023} developed a PoC system for the US Army Command (C5ISR), enabling cyberspace visualization and understanding. The architecture features components for data transmission and integration of tools that build a graph knowledge base from incoming data streams, supporting multi-level data correlation. The system thoroughly implements Cyber Situational Understanding (Cyber SU) capabilities required for tactical mission support. However, as a PoC developed by the US Command, it is not publicly available.

\vspace*{-0.1cm}
\subsection{Proposal of Dillabaugh et al.}

Dillabaugh et al.~\cite{dillabaugh2020cybercop} developed a PoC that enhances SA using a three-layer cyber-physical model. This system represents entities with nodes and employs simplified NATO Joint Military Symbology. The software, which was tested in defensive, offensive, and remediation scenarios, provides a detailed user guide and scenario explanations. However, it lacks flexibility for scenario modification and has limited data input options. 

\vspace*{-0.1cm}
\subsection{Proposal of Husak et al.}
 Husak et al.~\cite{HUSAK2022, Husak2024} performed CSA in heterogeneous and complex environments related to network infrastructure. Seeking to cover the entire OODA loop (Observe, Orient, Decide, Act) through use cases, where we see as the only paper in which the decomposition of missions into objectives is developed through the capabilities related to the Decide phase, and where it is possible to interact with the various assets either to mitigate or obtain feedback through the functionality related to the Act.



\vspace*{-0.1cm}
\subsection{Analysis of comparative table}
After analyzing the 5 CSA platform proposals in the comparative Table~\ref{tab:proposals-comp}, an analysis has been carried out. In particular, most platforms target technical or technical-strategic teams, with Esteve's solution addressing all users across levels~\cite{esteve2016cyber}. Data sources are predominantly internal, though some incorporate external or adversarial data. Although all platforms operate in cybernetic (CIS) environments, only Noel's solution extends to tactical military scenarios~\cite{Noel2023}. Common use cases include visualization of cyber knowledge graphs, mission impact assessment, and operational node visualization. All platforms prioritize visual representation through main maps or graph visualizations, with some offering advanced techniques like 3D modeling and virtual reality (Esteve). Military-specific features vary from NATO symbol integration (Dillabaugh) to mission dependency modeling (Noel) and decision support systems (Husak). Each platform demonstrates unique strengths: Kookjin in georeferencing and traffic visualization; Esteve offers immersive 3D and VR visualization; Noel provides comprehensive mission impact assessment; Dillabaugh emphasizes operational node visualization and mission engineering; while Husak focuses on incident handling and decision support. This comparison highlights the diverse approaches to CSA, with each platform tailoring its capabilities to specific operational needs and user groups.\\\\

The analyzed works show how CSA research focuses primarily on the first two levels of the SA process. For the civilian industry, available solutions such as Security Information and Event Management (SIEM, centered around aggregation and monitoring) are sufficient to improve the CSA level, since an industrial network is usually a homogeneous environment where only the cyber landscape is considered. More advanced solutions, such as SOAR, significantly enhance hardening capabilities and action execution within the network, aligning more closely with the ``Act'' category of the OODA loop.

Nevertheless, there is a notable lack of proposals or research on tools with active assistance capabilities for projection-related tasks, such as threat spread projection, mission decisive-condition identification, situation prediction, and impact analysis. This is the critical point for any NATO Federated Mission Networking service: having interoperability capabilities to aggregate information from C2ISR COI-services~\cite{natoC3baseline6} regardless of the domain (cyber or physical) and being able to infer information that assists decision-making profiles.



%% file: Proposal.tex
\section{Design of a military smart CSA platform}\label{sec:proposal} 

This section describes the 6 proposed criteria to be considered for the design of a military smart CSA platform, and then it introduces a high level architecture of platform built under such criteria.

We propose the concept of a military smart CSA platform to define a solution used in cyber defense operations, which leverages current market technologies to assist in data inference and decision-making. Such a platform should be able to provide a dynamic representation of the environment, both in terms of visual representation and managing the data plane, serving a wide array of user profiles and use cases, matching the following 3 main objectives:
\begin{itemize}
    \item Multi-dimension: Operation over different domains, integrating both data collection and interpretation capabilities as well as the ability to generate new information for situation prediction and action execution over available assets across the different dimensions managed by the CSA platform.
    \item Relevant information according to user profile: Information selectively presented and gathered based on their relevance to different user profiles in a given situation within a specific time frame. This characteristic ensures a platform applicable to different scenarios and usable by technical, operational, and strategic profiles.
    \item Multi-domain: Aggregated information from diverse data sources, establishing relationships between the cyber and kinetic domains (land, sea, air, space), enabling hybridized outputs that facilitate technical-operational analysis and the projection of actions across the different dimensions.
    
\end{itemize}

A previous work of Georgi et al.~\cite{Georgi2024} proposed quantitative evaluation criteria of CSA techniques and applications based on some aspects such as: timeliness, accuracy, and analyst's workload. Such mentioned aspects were obtained from the application of evaluation methodologies centered around the observation and interrogation of analysts, during and after a cyber-exercise, while using a CSA tool. However, in our case, we want to move away from a benchmark based on the tool's performance and aim to identify qualitative criteria related to capabilities that imply a cognitive or tactical advantage in a military scenario. Thus, the previous 3 main features associated with a military smart CSA platform spawn the following 6 criteria.

\subsection{Criteria 1: Multi-users adaptability}

Decision-making teams in civil cyber environments and hybrid military environments with tactical-strategic needs consist of diverse user profiles, e.g., tactical units, technicians, officers, and analysts. Thus, to maximize useful information delivery for each profile, a military smart CSA platform should adapt its interfaces and datasets to the user's expertise, avoiding the provision of irrelevant information.


\subsection{Criteria 2: Multiple quality data sources}

Most existing CSA platforms follow a data model that, at a minimum, fully covers cyber data ingestion and processing, e.g., topology analysis, identification of vulnerabilities in digital systems, and traffic analysis. Thus, a military smart CSA platform should extend their coverage to a broader range of quality data sources, e.g., data collection from SIEMs or SCADAs in industrial centers, interfaces with military C2 systems, and third-party datasets for analytical models. This variety of sources directly influences the tool's ability to operate across multiple physical dimensions by integrating data from diverse domains.

\subsection{Criteria 3: Dual operational environments}

CSA tools may operate in the cyber domain, showing certain hybrid capabilities such as the prediction of degradation of a tactical asset due to a compromise in a managing digital node. This capability includes, for example, a situation where the surveillance capabilities of a drone system will be affected if the controlling swarm system suffers a performance problem. In addition, CSA tools may operate in the Mission domain, highly focused on supporting tactical-operational planning of military operations, having as main capabilities the visualization of operational nodes (facilities or units), the mission course analysis, and the analysis of communications effectiveness. Thus, a military smart CSA platform should support both Cybernetic (CIS) and Mission (MI) operational environments.

\subsection{Criteria 4: CSA-level use cases}

Minimal covered use cases involve data gathering derived from network and asset audit and its posterior visualization, e.g., network traffic monitoring and digital asset inventory. Such minimal covered use cases are aligned with the first 2 CSA levels: perception and comprehension. However, more innovative tools provide the third SA level (projection), e.g., decision support via situation prediction or high-value target identification. The projection level is strongly related to the final phase of the OODA loop, i.e., Act, which enables actions such as automatic configuration of firewalls or automatic shutdown of systems to avoid threat propagation. Thus, a military smart CSA platform should support use cases for the 3 CSA levels: perception, comprehension, and projection.

\subsection{Criteria 5: Enriched visualization}

Some works, e.g., ~\cite{Jiang2022}, reveal limitations of CSA platforms in terms of visualization capabilities, which directly impact situational understanding. Even if it can be difficult to establish a full taxonomy of visualizations, it is possible to identify that some novel visualizations ease the comprehension of information and boost human inference. Thus, a military smart CSA platform should support minimal visualization techniques based on: knowledge graphs, C2-cybernetic hybridized views, and georeferenced views.

\subsection{Criteria 6: Military approach}

Finding references about CSA platforms for use in military environments may be difficult because of their confidentiality. However, some non-confidential works~\cite{esteve2016cyber, Machado2014ConceptualAF} show that CSA platforms are widely used in defense. Thus, a military smart CSA platform should support different use cases that are quite specific for the defense sector and probably are not found in other sectors, e.g., visualization of operational nodes network and identification of high-value targets, compilation of a military-industrial COP, mission projection, and risk analysis on tactical assets, among others.

\begin{figure*}[h]
\centering
	\includegraphics[width=0.9\textwidth]{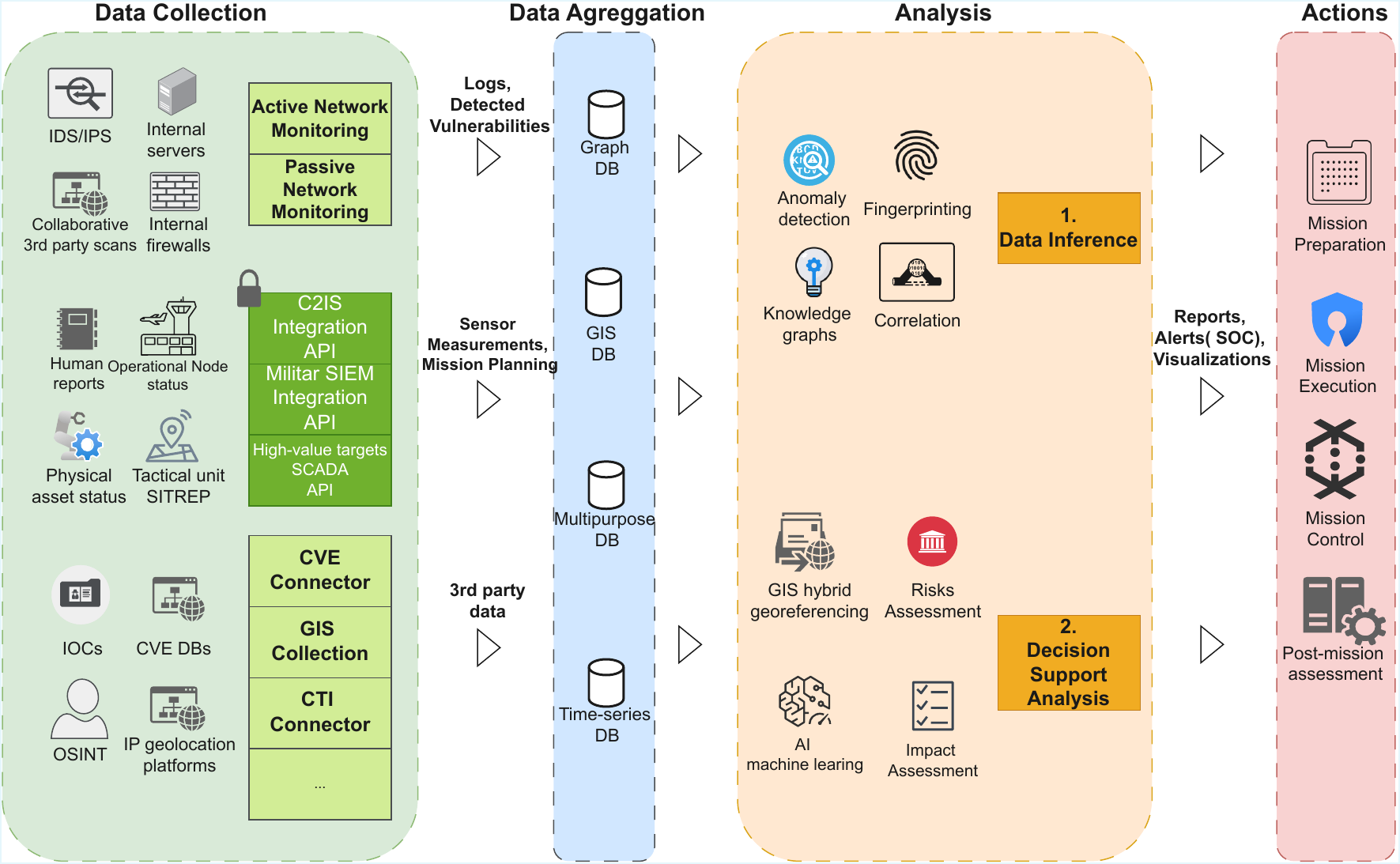} 
	\caption{High level architecture of a Smart CSA platform}
	\label{fig:CSAArchitecture}
\end{figure*}

Throughout the study, numerous civilian and military architectural proposals have been examined. Still, the absence of a common architecture has led us to create one for a hybrid CSA platform applicable to both fields. Therefore, in Figure~\ref{fig:CSAArchitecture}, it is possible to see the proposed design.

This design consists of 4 basic components identified as fundamental for a CSA platform. The first is responsible for collecting and sending information to the second module, where data aggregation occurs. After data correlation, the analysis module processes the information and generates a series of options that facilitate decision-making. Finally, this data is sent to the action module responsible for decision-making.

%% file: Experiments.tex
\section{Experiments}\label{sec:experiments} 

This section develops the validation of the 6 proposed criteria for a military smart CSA platform. We will use a real open-source toolset as a testbed to simulate a cyber operation, observe its performance, and evaluate the degree to which each criterion is met, as well as how this affects overall CSA.

In the scientific literature used as a reference, we observe well-defined research trends, i.e., studies mainly focus on conceptually demonstrating the applicability of CSA in cyberwarfare and its short-term importance. However, a few proposals go beyond the theoretical framework and use current market technologies to shed light on how CSA tools impact real-world scenarios in operational-strategic decision-making.

Our experiment will consist of deploying and testing the most mature tool among the five proposals studied in Section~\ref{sec:comparison}. The selected tool is developed from the work of Husák et al., the CRUSOE toolset. The reason for selecting CRUSOE is that it is a tool whose background, early development stages, final product (and later extensions) have been documented~\cite{Husak2021,HUSAK2022,Husak2024} over four years in various scientific publications. This tool was developed by the CSIRT-MU (Computer Security Incident Response Team of Masaryk University) and evaluated within the Masaryk University network for three months, creating a knowledge graph of 697,783 nodes and 22,119,299 edges with detailed information about 29,535 unique IP addresses in the network~\cite{HUSAK2022}.

This toolset is open-source and finely crafted to be based around the OODA loop, featuring the following capabilities:
\begin{itemize}
    \item Continuous data aggregation (active and passive network monitoring, third-party sourcing, local data sourcing).
    \item Visualization dashboard.
    \item Mission decomposition and mission resilience calculation.
    \item Search engine that identifies an asset's similar nodes in the knowledge graph.
    \item Threat spread projection graph.
    \item Integration with network security tools APIs to execute actions (firewall, DNS, mail filters, Active Directory, etc.).
\end{itemize}

\begin{figure}[!h]
\centering
	\includegraphics[width=\columnwidth]{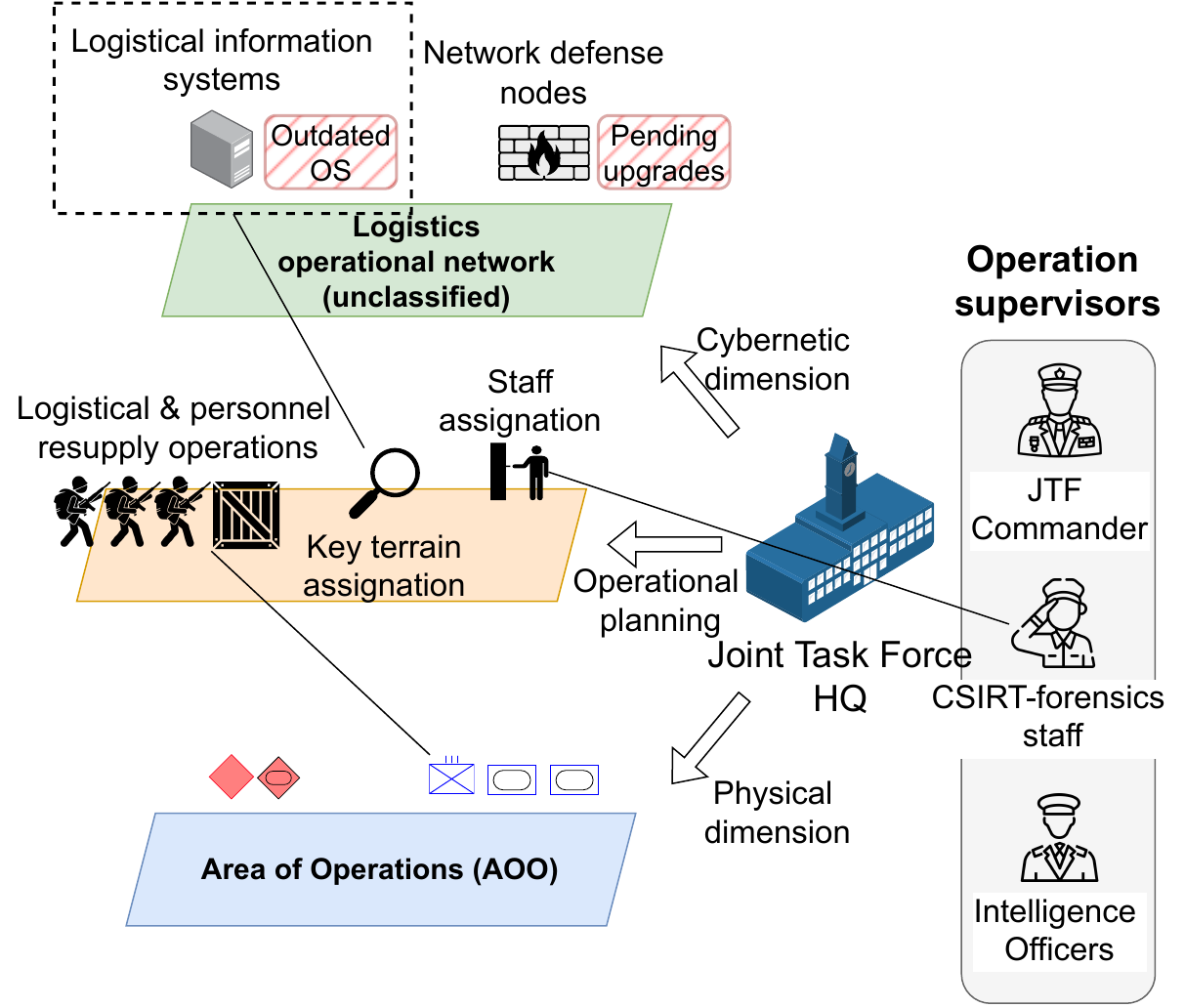} 
	\caption{Graphic representation of Dressler~\cite{Dressler2014} CSA use case}
	\label{fig:DresslerUC}
\end{figure}

The validation will be conducted considering a specific cyber defense scenario: an adaptation of the scenario described in the work of Dressler et al.~\cite{Dressler2014}, which considers a military logistical operation where kinetic-cybernetic cognitive capabilities must be deployed to fuse the threat surface dimensions into a single plane that allows identification of admissible mission trade-offs and success conditions. This hypothetical case is graphically represented in Figure~\ref{fig:DresslerUC}. A U.S. Joint Task Force (JTF) heavily relies on a logistical-support information system designed to maintain personnel and supply logistics during combat operations, strategically defining this system as cyber key terrain. The supervision of these operations involves personnel with various roles: JTF commander, intelligence officers, and a cyber-forensics team.

The narrative of the use case presents allied forces with a series of conditions that deteriorate the defensive capability of the logistical network: intelligence reports indicating a change in the adversary's prioritized target list, and the discovery of a critical vulnerability in the logistical-support system. This scenario presents a set of variables that positively and negatively affect the capabilities of allied forces across different AOO dimensions and all phases of the OODA loop (Observe, Orient, Decide, Act): operational readiness of the cyber-forensics team, applicability of mitigation measures, intelligence reports provided by personnel, etc.



The experiments were conducted by deploying CRUSOE using Docker on a single server (4-core Intel N100 CPU, 8GB RAM), and the datasets used in the experiment were obtained from the official CRUSOE repository~\cite{CRUSOEog}. These datasets were loaded into CRUSOE's core database, Neo4j. Next, we describe the evaluation results of the six criteria proposed in Section~\ref{sec:proposal} for CRUSOE.

\subsection{Criteria 1: Multi-users adaptability}

CRUSOE's different modules achieve a certain level of profile adaptation. Its dashboard for monitoring all information is tailored for incident response teams that need an overview of the cybersecurity posture. Meanwhile, the Decision and Act functionalities have an interface designed for network operators involved in strategic planning but maintain an interactive format, easing the cognitive workload for technical experts and keeping a partially friendly UI for decision-makers.

CRUSOE partially meets the criteria for user adaptation in a smart CSA Platform. Its modular design effectively supports different user profiles but remains primarily framed within the technical profile. There is no user-driven customization, such as dynamic interfaces that adapt based on expertise level or role-based information filtering.

\subsection{Criteria 2: Multiple quality data sources}

CRUSOE provides extensive cybersecurity monitoring and decision-support capabilities as it aggregates data from multiple cybersecurity sources, such as public vulnerability databases (e.g., NVD, OSVDB, Exploit Database) and manufacturer-specific vulnerability feeds (e.g., Cisco, Microsoft). Theoretically, it also combines passive network monitoring (via NetFlow/IPFIX) with active scanning (Nmap) to detect vulnerabilities in networked assets.

However, during the real use of CRUSOE, we observed limited active monitoring capabilities, as most data sourcing focused on the preloaded information. Indeed, a partial analysis of the real-world deployment was performed, as information from the physical host where the platform was deployed was detected and added to the Graph Representation DB, but no further scan capabilities were observed. 

\subsection{Criteria 3: Dual operational environments}

The CRUSOE platform is primarily designed for cybersecurity situational awareness rather than tactical-operational military planning. Its architecture and functionalities focus on network security monitoring, vulnerability analysis, and automated response.

Although the Act capability of the CRUSOE is tightly coupled with security nodes, such as firewalls and switches, within the topology of the fictional scenario, there has been no real test of the performance of configuration profiles applied to network nodes due to the absence of a configurable network topology. This aligns with our statement about CRUSOE being more of a cyber-focused CSA tool than allowing mission planning and multi-domain operations. One example is the lack of kinetic dimension features, such as operator mission assignment, a commonly proposed capability in other works, and very useful for the selected use case.

\subsection{Criteria 4: CSA-level use cases}
The key element to choose CRUOSE as the starting target to benchmark is that it features at least one capability framed within each phase of the OODA loop. The scientific work and manuals \cite{HUSAK2022, Husak2024, CRUSOEog} cite data aggregation based on key data sources (Observe), a well-fitted cyber data-model that allows the construction of knowledge graphs and ease the representation of data (Orient), a set of capabilities that allow ``mission decomposition'', ``mission resilience calculation'', ``prediction of threat spreads'', ``data correlation optimized search-engine'' (Decide) and finally multiple network services integrations with CRUSOE e.g., DNS, firewalls, mail- to allow for application of security profiles based on the infered information from the Decide phase's data.  

As a matter of fact, our evaluation of this criterion using the toolset is not yet completed. Decision-assistance capabilities were not fully tested as the deployment of the complete toolset was not doable due to technical complications. Due to this same reason, only ``missions decomposition'' and ``prediction of threat spread'' have been tested regarding the Decision phase's capabilities. Still, the enabled subset of Act capabilities, along with the complete Observe capabilities, provides a cognitive value high enough to allow the users to visualize the cyber-environment and easily identify the attack surface of the high-value services identified by the Network Scan Data, enhancing the overall level of CSA of the operator.

\subsection{Criteria 5: Enriched visualization}

CRUSOE utilizes network graphs, dashboards, and attack modeling to help analysts identify security risks. However, its visualization capabilities are primarily limited to cyber network representations and do not fully extend into hybridized views that integrate cyber and kinetic domains.

Most visualization work is done through CRUSOE's dashboard (Orient module). While CRUSOE effectively represents cyber data, it lacks more advanced visualization models such as geospatial overlays, heatmaps, or interactive hybrid dashboards. Its primarily graph-oriented dashboard visualization is also not easily navigable, as in some cases, access to the graph-representation DB is necessary to study omitted needed to better understand the context of the knowledge graph presented in the main dashboard.

\subsection{Criteria 6: Military approach}

CRUSOE provides strong capabilities in network monitoring, risk assessment, and automated response. However, its alignment with military-specific use cases appears to be limited based on the documentation and conducted benchmark. On paper, it features different APIs that allow CRUSOE to interact with external services such as firewall configurations, DNS, and mail filters, suggesting the possibility of developing APIs to integrate SCADA or C2 systems.

Currently, the most attractive feature for military cyber defense is the Decide capabilities. This capability uses attack graphs and Bayesian inference to predict security breaches, a method that could potentially be adapted to military risk analysis on tactical assets.




%% file: Conclusions.tex
\section{Conclusions and Future Work}\label{sec:conclusions} 

In this paper, we have compared different CSA platforms. This study has been done through the search of different proposals and studies carried out, where different architectures are proposed, what functionality they offer, and whether they are conceptual studies or if the tool has been developed. To make this comparison, a table was created considering a series of proposed criteria. These criteria have been presented to support future developments and used to evaluate the different studies. Finally, the CRUSOE platform~\cite{CRUSOE} has been used to test the criteria defined in this tool to measure its degree of maturity and identify expansion opportunities.  

CRUSOE has several limitations when used in a military-grade CSA platform, due it is designed primarily for civilian network management. It lacks hierarchical data and decision processing, military classification levels for chain-of-command protocols, and military-oriented features in the ACT component, including geospatial visualization and tools for targeted attack analysis. Additionally, its reliance on Neo4j results in high costs for military use, and it is not equipped for multi-domain or multinational joint operations due to scalability issues. To improve CRUSOE's suitability for military applications, enhancements are needed in the Decide and Act phases, which currently demand the most research. This involves expanding its capabilities to operate in dual-operation enviroment by integrating tools to manage complex assets with dependencies across various dimensions.

Future work will explore the specialization of the CRUSOE platform as a CSA tool that can be effectively employed by military commanders. Additionally, we have proposed a set of qualitative criteria to evaluate and develop SA tools, enabling assessment of the tactical and cognitive advantages these tools provide in military scenarios.

%% file: bibliography.bib
@misc{EDF,
  title = {{European Defense Fund} Work Programme 2025},
  howpublished = {\url{https://defence-industry-space.ec.europa.eu/edf-work-programme-2025_en}},
  note = {Accessed: 2025-03-20}
}

@article{Endsley1995,
author = {Mica R. Endsley},
title ={Toward a Theory of Situation Awareness in Dynamic Systems},
journal = {Human Factors},
volume = {37},
number = {1},
pages = {32-64},
year = {1995},
doi = {10.1518/001872095779049543},
URL = { 
    
        https://doi.org/10.1518/001872095779049543
    
    

},
eprint = { 
    
        https://doi.org/10.1518/001872095779049543
    
    

}
}

@article{FRANKE2014,
title = {Cyber situational awareness – A systematic review of the literature},
journal = {Computers \& Security},
volume = {46},
pages = {18-31},
year = {2014},
issn = {0167-4048},
doi = {https://doi.org/10.1016/j.cose.2014.06.008},
url = {https://www.sciencedirect.com/science/article/pii/S0167404814001011},
author = {Ulrik Franke and Joel Brynielsson}
}

@article{Jiang2022,
author = {Jiang, Liuyue and Jayatilaka, Asangi and Nasim, Mehwish and Grobler, Marthie and Zahedi, Mansooreh and Ali Babar, Muhammad},
year = {2022},
month = {01},
pages = {1-1},
title = {Systematic Literature Review on Cyber Situational Awareness Visualizations},
volume = {10},
journal = {IEEE Access},
doi = {10.1109/ACCESS.2022.3178195}
}

@inproceedings{Georgi2024,
author = {Nikolov, Georgi and Perez, Axelle and Mees, Wim},
title = {Evaluation of Cyber Situation Awareness - Theory, Techniques and Applications},
year = {2024},
isbn = {9798400717185},
publisher = {Association for Computing Machinery},
address = {New York, NY, USA},
url = {https://doi.org/10.1145/3664476.3670921},
doi = {10.1145/3664476.3670921},
booktitle = {Proceedings of the 19th International Conference on Availability, Reliability and Security},
articleno = {70},
numpages = {10},
location = {Vienna, Austria},
series = {ARES '24}
}

@inproceedings{Machado2014ConceptualAF,
  title={Conceptual Architecture for Obtaining Cyber Situational Awareness},
  author={Andre Ferreira Alves Machado and Edgar Toshiro Yano},
  year={2014},
  url={https://api.semanticscholar.org/CorpusID:108175837}
}

@article{Skopik2022,
author = {Skopik, Florian and Bonitz, Arndt and Grantz, Volker and Göhler, Günter},
year = {2022},
month = {09},
pages = {1-25},
title = {From scattered data to actionable knowledge: flexible cyber security reporting in the military domain},
volume = {21},
journal = {International Journal of Information Security},
doi = {10.1007/s10207-022-00613-7}
}

@INPROCEEDINGS{Llopis2018,
  author={Llopis, Salvador and Hingant, Javier and Pérez, Israel and Esteve, Manuel and Carvajal, Federico and Mees, Wim and Debatty, Thibault},
  booktitle={2018 International Conference on Military Communications and Information Systems (ICMCIS)}, 
  title={A comparative analysis of visualisation techniques to achieve cyber situational awareness in the military}, 
  year={2018},
  volume={},
  number={},
  pages={1-7},
  doi={10.1109/ICMCIS.2018.8398693}}

@article{SOTELOMONGE2021869,
title = {Conceptualization and cases of study on cyber operations against the sustainability of the tactical edge},
journal = {Future Generation Computer Systems},
volume = {125},
pages = {869-890},
year = {2021},
issn = {0167-739X},
doi = {https://doi.org/10.1016/j.future.2021.07.016},
url = {https://www.sciencedirect.com/science/article/pii/S0167739X21002788},
author = {Marco Antonio {Sotelo Monge} and Jorge {Maestre Vidal}}
}

@inproceedings{Rodríguez-Bermejo2021,
author = {Sandoval Rodr\'{\i}guez-Bermejo, David and Maestre Vidal, Jorge and Est\'{e}vez Tapiador, Juan},
title = {The Stress as Adversarial Factor for Cyber Decision Making},
year = {2021},
isbn = {9781450390514},
publisher = {Association for Computing Machinery},
address = {New York, NY, USA},
url = {https://doi.org/10.1145/3465481.3470047},
doi = {10.1145/3465481.3470047},
booktitle = {Proceedings of the 16th International Conference on Availability, Reliability and Security},
articleno = {101},
numpages = {10},
location = {Vienna, Austria},
series = {ARES '21}
}

@article{CyCOP2023,
  author = {Kookjin Kim and Jaepil Youn and Sukjoon Yoon and Jiwon Kang and Kyungshin Kim and Dongkyoo Shin},
  title = {Study on Cyber Common Operational Picture Framework for Cyber Situational Awareness},
  journal = {Applied Sciences},
  volume = {13},
  number = {4},
  pages = {2331},
  year = {2023},
  doi = {10.3390/app13042331}
}

@article{esteve2016cyber,
  title={Cyber Common Operational Picture: A Tool for Cyber Hybrid Situational Awareness Improvement},
  author={Esteve, Manuel and P{\'e}rez, Israel and Palau, Carlos and Carvajal, Federico and Hingant, Javier and Fresneda, Miguel A and Sierra, Juan P},
  journal={North Atlantic Treaty Organization (NATO) Science and Technology Organization (STO), Technical Report STO-MP-IST-148},
  year={2016}
}

@article{HUSAK2022,
title = {CRUSOE: A toolset for cyber situational awareness and decision support in incident handling},
journal = {Computers \& Security},
volume = {115},
pages = {102609},
year = {2022},
issn = {0167-4048},
doi = {https://doi.org/10.1016/j.cose.2022.102609},
url = {https://www.sciencedirect.com/science/article/pii/S0167404822000086},
author = {Martin Husák and Lukáš Sadlek and Stanislav Špaček and Martin Laštovička and Michal Javorník and Jana Komárková}
}

@INPROCEEDINGS{Husak2024,
  author={Husák, Martin and Sadlek, Lukáš and Hesko, Martin and Šebela, Vít and Špaček, Stanislav},
  booktitle={2024 20th International Conference on Network and Service Management (CNSM)}, 
  title={The Evolution of the CRUSOE Toolset: Enhancing Decision Support in Network Security Management}, 
  year={2024},
  volume={},
  number={},
  pages={1-3},
  doi={10.23919/CNSM62983.2024.10814288}
}

@article{dillabaugh2020cybercop,
  title={CyberCOP: Cyber Situational Awareness Demonstration Tool},
  author={Dillabaugh, C and Bennett, D},
  journal={Defence Research and Development Canada, Ottawa, Canada},
  year={2020}
}

@article{Noel2023,
author = {Steven Noel and Stephen Purdy and Annie O’Rourke and Edward Overly and Brianna Chen and Christine DiFonzo and Joseph Chen and George Sakellis and Mandira Hegde and Mano Sapra and Corrine Araki and Jeremy Martin and Ben Koehler and John Keenan and Timothy Coen and William W Watson and Jerry Harper and Kevin Jacobs},
title ={Graph analytics and visualization for cyber situational understanding},
journal = {The Journal of Defense Modeling and Simulation},
volume = {20},
number = {1},
pages = {81-95},
year = {2023},
doi = {10.1177/15485129211051385},
URL = {https://doi.org/10.1177/15485129211051385},
eprint = {https://doi.org/10.1177/15485129211051385}
}

@ARTICLE{Torvald2023,
AUTHOR={Ask, Torvald F.  and Knox, Benjamin J.  and Lugo, Ricardo G.  and Hoffmann, Lukas  and Sütterlin, Stefan },
TITLE={Gamification as a neuroergonomic approach to improving interpersonal situational awareness in cyber defense},
JOURNAL={Frontiers in Education},
VOLUME={8},
YEAR={2023},
URL={https://www.frontiersin.org/journals/education/articles/10.3389/feduc.2023.988043},
DOI={10.3389/feduc.2023.988043},
ISSN={2504-284X}}

@INPROCEEDINGS{Dressler2014,
  author={Dressler, Judson and Bowen, Calvert L. and Moody, William and Koepke, Jason},
  booktitle={2014 6th International Conference On Cyber Conflict (CyCon 2014)}, 
  title={Operational data classes for establishing situational awareness in cyberspace}, 
  year={2014},
  volume={},
  number={},
  pages={175-186},
  doi={10.1109/CYCON.2014.6916402}
}

@techreport{natoC3baseline6,
  title        = {C3 Taxonomy Baseline 6},
  author       = {{NATO Communications and Information Agency}},
  institution  = {NATO C3 Board},
  year         = {2023},
  month        = {October},
  address      = {Brussels, Belgium},
  note         = {Available at: \url{https://www.nato.int/cps/en/natohq/topics_157573.htm}},
  type         = {Technical Report}
}

@misc{CRUSOEog,
  title = {{original CSIRT-MU CRUSOE},
  howpublished = {\url{https://github.com/CSIRT-MU/CRUSOE}},
  note = {Accessed: 2025-03-21}
}}

@misc{CRUSOE,
  title = {{UMU CRUSOE testing environment},
  howpublished = {\url{https://github.com/AnthonyFA/CDLCRUSOE/tree/dev}},
  note = {Accessed: 2025-03-21}
}}

@inproceedings{Husak2021,
author = {Hus\'{a}k, Martin and La\v{s}tovi\v{c}ka, Martin and Tovar\v{n}\'{a}k, Daniel},
title = {System for Continuous Collection of Contextual Information for Network Security Management and Incident Handling},
year = {2021},
isbn = {9781450390514},
publisher = {Association for Computing Machinery},
address = {New York, NY, USA},
url = {https://doi.org/10.1145/3465481.3470037},
doi = {10.1145/3465481.3470037},
booktitle = {Proceedings of the 16th International Conference on Availability, Reliability and Security},
articleno = {112},
numpages = {8},
location = {Vienna, Austria},
series = {ARES '21}
}
